\documentclass{iau}

\usepackage{amsmath}
\usepackage{graphicx}
\usepackage{multirow}

\begin{document}

\lefttitle{Sallmen \& Korpela}
\righttitle{Exploring Orbital Stability of Large, Lightweight Mirrors around Exoplanets}

\jnlPage{1}{7}
\jnlDoiYr{2026}
\doival{10.1017/xxxxx}

\aopheadtitle{Proceedings IAU Symposium}
\editors{J. Haqq-Misra \&  R. Kopparapu, eds.}

\title{Exploring the Orbital Stability of Large, Lightweight Mirrors around Exoplanets}

\author{Shauna M. Sallmen$^1$ \& Eric J. Korpela$^2$}
\affiliation{$^1$University of Wisconsin - La Crosse, La Crosse, WI, USA}
\affiliation{$^2$Space Sciences Laboratory at the University of California, Berkeley, CA, USA}

\begin{abstract}
Extraterrestrial civilizations might place large, lightweight mirrors into orbit around an exoplanet, either to alter its climate or to provide illumination to the planet’s dark side. We previously analyzed the detectability of a fleet of 1 km x 1 km, 1000 kg mirrors \citep{ksg2015}. Because these mirrors are large and lightweight, their orbits are significantly affected by the star’s radiation pressure (RP). We created a simulation package based on the REBOUND N-body integrator, incorporating RP that directs starlight towards the planet’s center. RP can always affect mirrors, or only during orbital night. We have simulated mirrors in initially circular orbits around exoplanets at various locations in the habitable zones of eight types of main sequence stars.  Initial mirror orbit sizes range from 2 to 10 planet radii, and we included 4 different initial orbit orientations. For each simulation, we have the mirror’s survival time, trajectory, distance from the planet at each time, and velocity relative to the planet at each time. We present an analysis of trends in mirror orbit stability, and relate these to the ratios of the RP and gravitational accelerations, as well as the ratio of planet orbit period to mirror orbit period. 
\end{abstract}

%\begin{keywords}
%Key1, Key2, Key3, Key4
%\end{keywords}

\maketitle

\section{Introduction}

\begin{figure}[!h]
\centering
   \includegraphics[width=7cm]{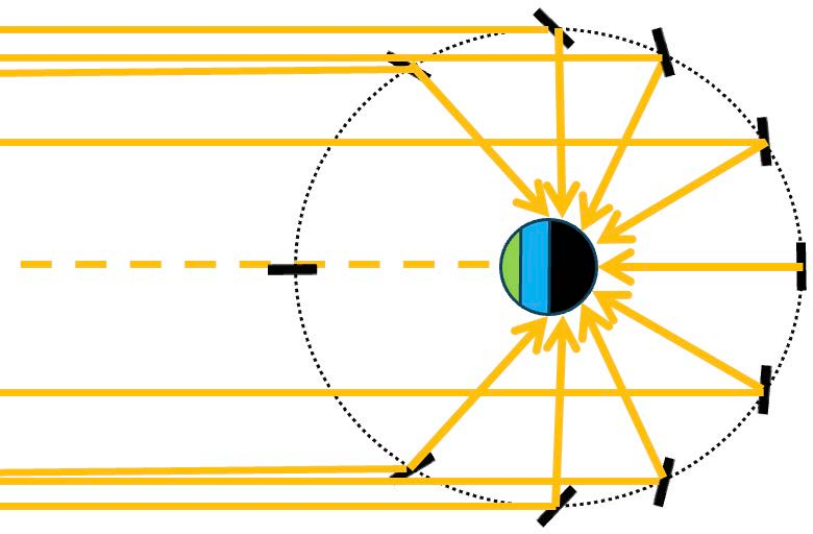}
   \caption{Schematic diagram of starlight redirected onto a planet using orbiting mirrors.}
    \label{fig:illuminate}
\end{figure}

Extraterrestrial societies might redirect starlight onto a planet using a fleet of large, lightweight, steerable mirrors, as illustrated in Figure~\ref{fig:illuminate}. Potential applications include terraforming a planet with unsuitable climate, or illuminating the dark side of a planet in synchronous rotation. 
  
For orbiting mirrors with a large surface area and low mass, radiation pressure (RP) will significantly affect the motion and orbit stability of these mirrors. Very unstable orbits likely require large amounts of fuel to maintain orbit.

In this work, we seek to identify mirror orbits that are more stable (and therefore more fuel efficient) in the presence of RP.

\section{Methods}

We simulated mirror motions with a Python package that uses the REBOUND N-body simulator \citep{ReinLiu2012} with the adaptive integrator IAS15 \citep{ReinSpiegel2015}. 

In each simulation, 
we placed an Earth-sized planet orbiting at the inside, middle, or outside edge of the Habitable Zone (HZ) of different main-sequence stars, spanning spectral types M8 to G2. Stellar properties are from \citet{zombeck}, while HZ boundaries were determined using optimistic values of \citet{kopparapuetal13}, accessed at \url{http://depts.washington.edu/naivpl/content/hz-calculator}. These star types were chosen because their potentially habitable exoplanets have a chance of tidal synchronization within 6.2 Gyr, using the methods and equation 3 of \citet{ForgetLeconte2013}. 

\begin{figure}[h]
 \centering
   \includegraphics[width=5.8cm]{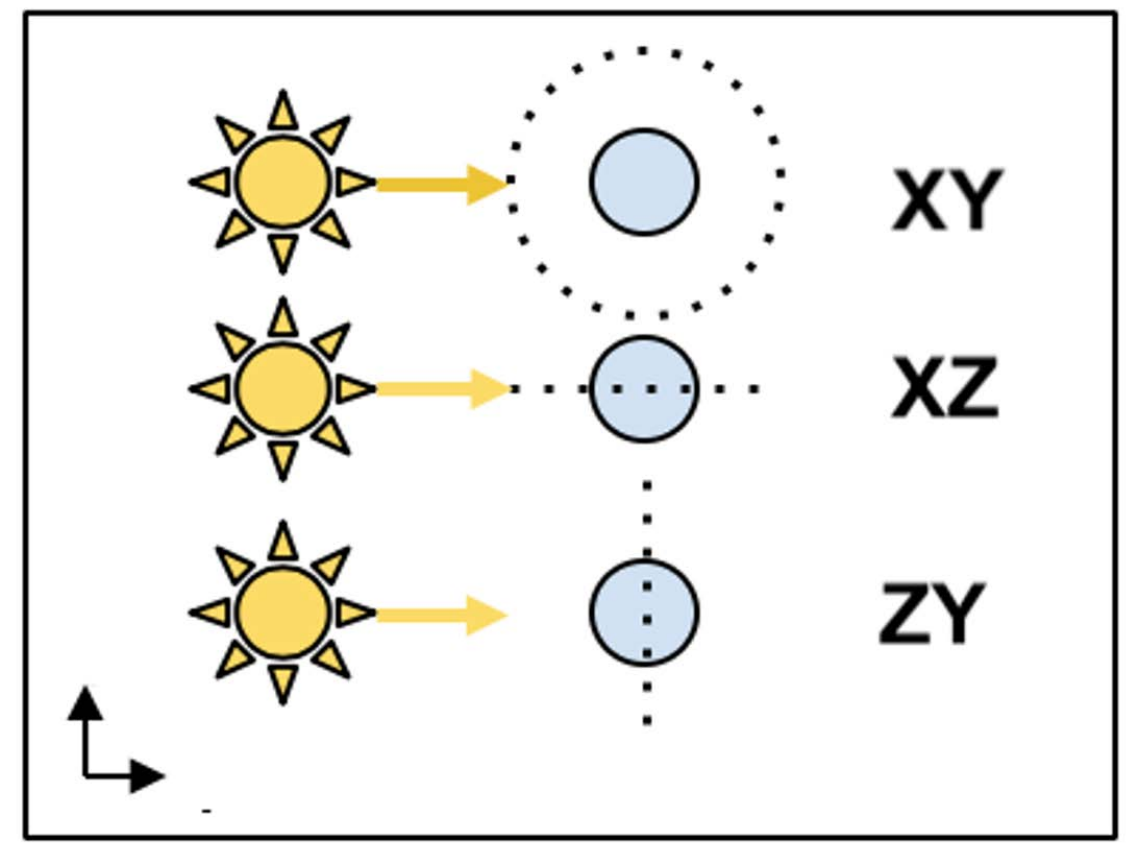}
   \caption{Schematic diagram of initial mirror orbit orientations.}
    \label{fig:orientation}
\end{figure}

Mirrors with a mass of 1000 kg and an area of 1 km$^2$ were initially placed in circular orbits at distances of 2, 3, and 10 planet radii, using four different mirror orbit orientations, as illustrated in Figure~\ref{fig:orientation}. The planet's orbit defined the XY plane, and initial mirror orbit orientations could be: (1) in the same plane and direction as the planet (+XY), (2) in the same plane but opposite the planet's orbital motion ($-$XY), (3) perpendicular to the planet's orbit and edge-on to the star (XZ), or (4) with an orbit face-on to the star along the planet's day-night line (ZY). 

Each planet-mirror configuration was simulated with (1) no RP (for comparison purposes), (2) RP always acting on the mirror, and (3) RP acting only when the mirror is on the night side of the planet. RP contributions to mirror accelerations were calculated from stellar luminosity ($L$), mirror area ($A$) and mass ($m$), planet distance from star ($d$), and angle of incidence $\theta$ using
\begin{equation}
    a_{\rm RP} = \frac{{2L A \cos^2\!{\theta}}}{4\pi c d m},
\end{equation}
assuming that the mirror is oriented so that starlight is redirected towards the planet center. Simulations were run for 1000 initial orbit periods, stopping early if the mirror crashed or escaped. 

Simulation output included the mirror orbit survival time, as well as relative positions and motions for the star, planet, and mirror 100 times per initial mirror orbit period. 

\section{Results \& Discussion}

\subsection{Trends for Planets at Inner Edge of HZ}

There were 216 simulations for planets located at the inner edge of the HZ, 72 for each RP possibility. In 22 of the simulations with no RP involving M stars, the mirror did not survive. In these simulations, the planet is close to its star, so stellar gravitational effects alone destabilize the mirror orbits. After removing the 44 equivalent RP simulations from the analysis, several trends (not rules) emerged.

\begin{figure}[h]
  \centering
   \includegraphics[width=6.3cm]{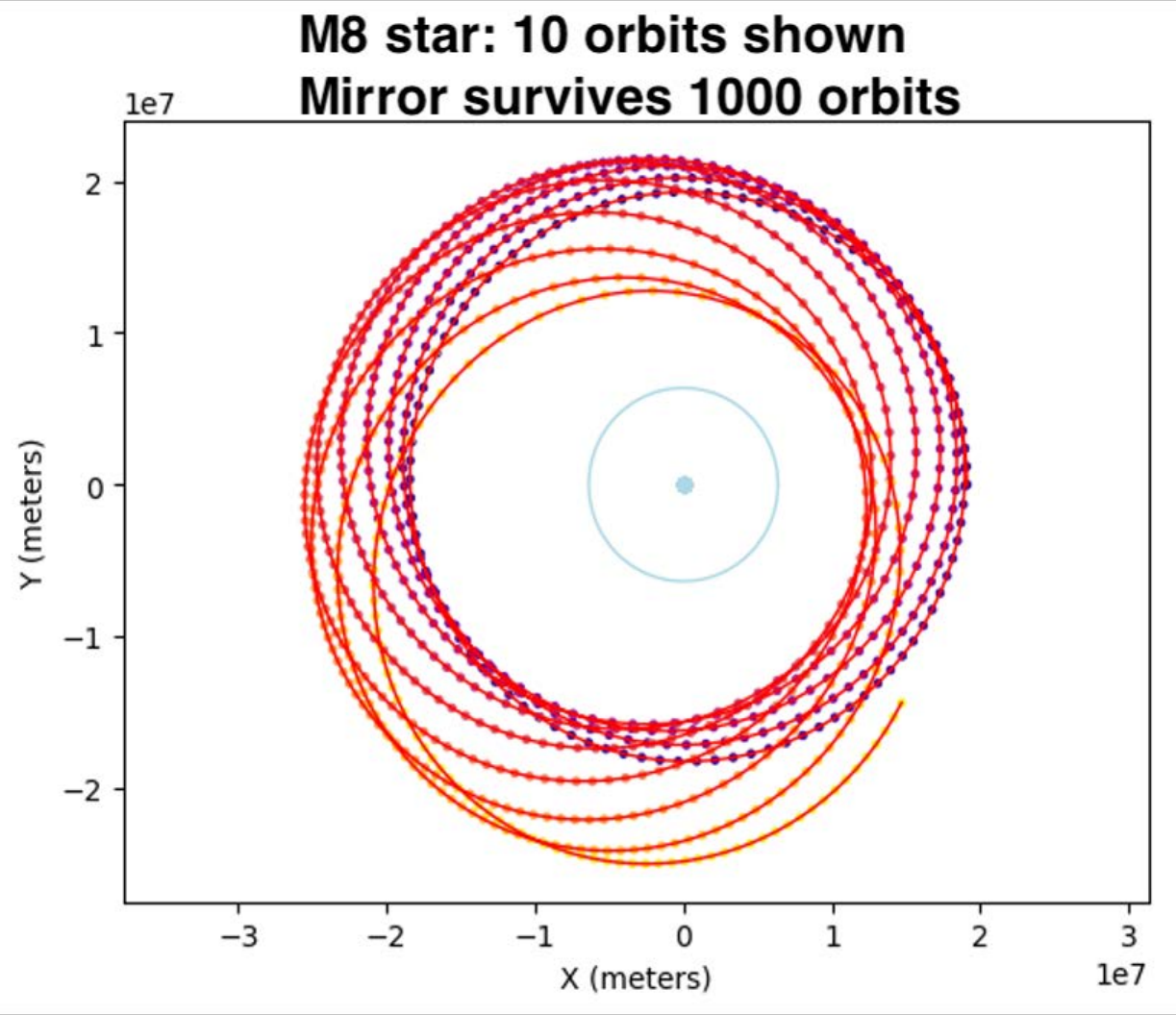}
   \caption{Part of the mirror orbit trajectory around its planet, for a mirror that survives a full 1000 orbits. 
    The inner dot represents the planet center, while the circle outline indicates its surface.}
    \label{fig:M_precession}
  \end{figure}

Mirrors around planets orbiting M5 and M8 stars were more likely to survive the full simulation than those orbiting more massive and luminous stars. In these situations, the HZ is very close to the star, so there are fewer mirror orbits per planet orbit and the RP direction changes rapidly. 
RP-induced precession aids mirror survival for some of these situations. 
This orbital precession is clearly visible in Figure~\ref{fig:M_precession}, which shows the partial trajectory of a mirror for one such simulation. In this case, the mirror always experienced RP and was initially placed in a +XY orbit 3 planet radii from the center of a planet orbiting an M8 star.

%\vspace{-0.2cm}

Mirrors with initially face-on (ZY) orbits were also more likely to survive, especially when always subject to radiation pressure and for more massive stars. In these simulations, the initial angle of incidence between starlight and mirrors is quite large, so the effects of RP are small. For more massive stars, the direction of starlight changes slowly, as the habitable zone is further out and there are many mirror orbits per planet year, so the RP effects remain small.

\begin{figure}[!h]
 \centering
   \includegraphics[width=13cm]{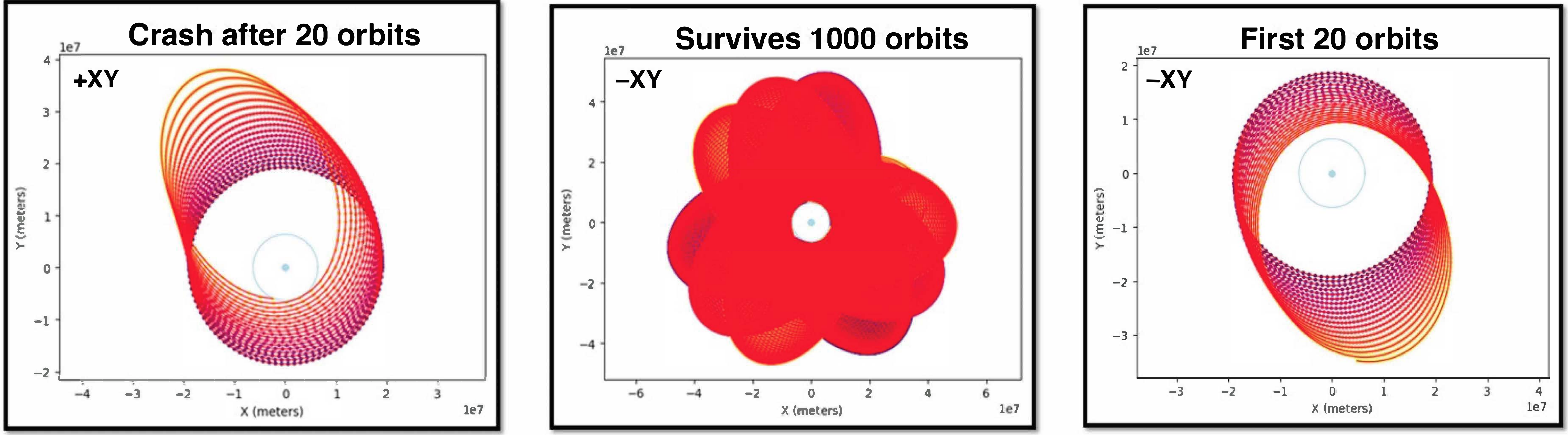}
   \caption{Comparison of +XY and $-$XY mirror trajectories around a planet orbiting an M0 star at its inner HZ edge, and experiencing RP only during orbital night. The mirror is initially at 3 Earth radii. }
    \label{fig:Reverse}
  \end{figure}

Mirrors orbiting opposite to the planet orbit direction ($-$XY) tend to survive longer than those orbiting in the same direction as the planet (+XY). Figure~\ref{fig:Reverse} illustrates one extreme example. The left panel shows the entire mirror orbit trajectory for the +XY orientation, ending when the mirror crashes into the planet. The middle panel shows the full 1000-orbit trajectory for the $-$XY simulation, while the right panel shows the first 20 orbits of the $-$XY simulation. The RP-induced mirror orbit elongation is smaller in the $-$XY case, so that RP-induced orbit precession allows mirror survival. These different survival times for the two opposing orbit orientations are likely due to differing
angular momentum transfer from the planet to its mirror. 

\subsection{Including additional Planet Locations}

\begin{figure}[h]
  \centering
   \includegraphics[width=11cm]{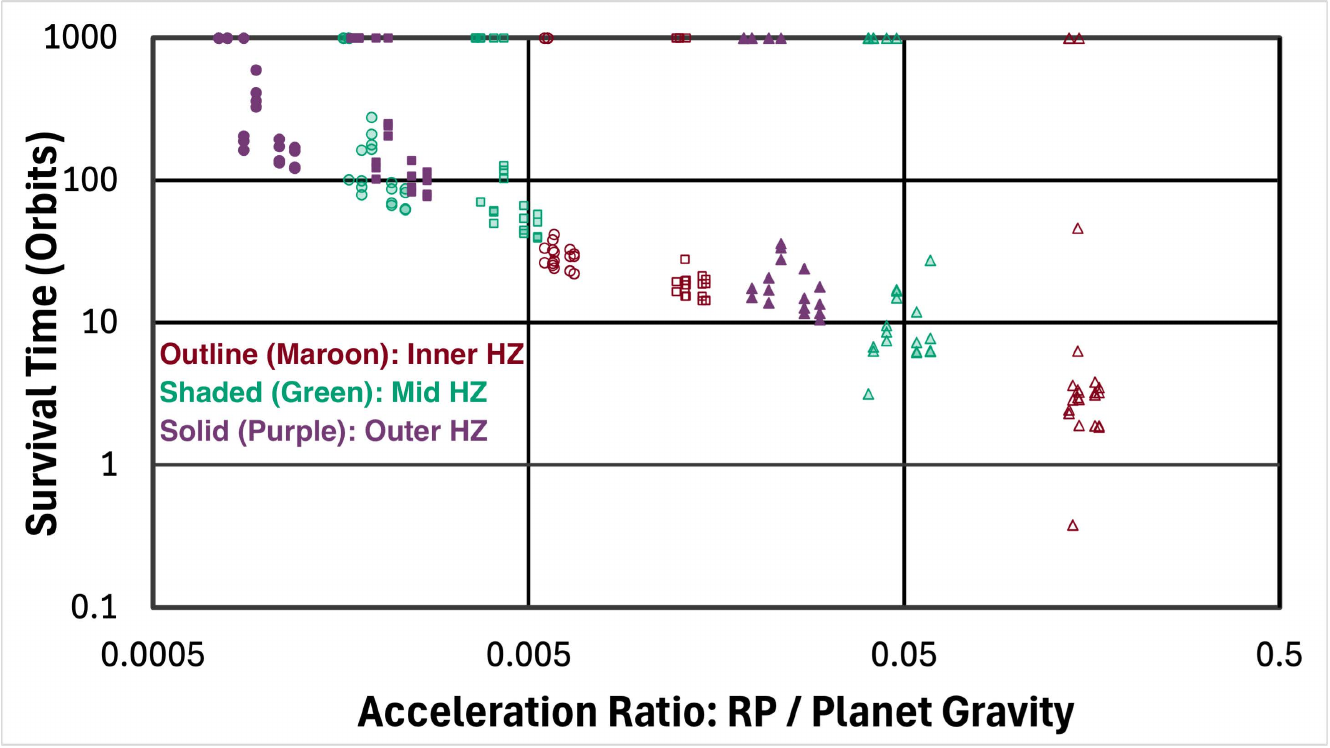}
   \caption{Mirror survival time in initial orbit periods for simulations with initial mirror orbits in the same plane as the planet. Circles, squares, and triangles are for mirrors initially at 2, 3, and 10 planet radii. 
   %Maroon: planet at inner HZ edge. Green: planet in middle of HZ. Purple: planet at outer HZ edge.
    Outline Symbols (Maroon): planet at inner HZ edge. Shaded Symbols (Green): planet in middle of HZ. Solid Symbols (Purple): planet at outer HZ edge.
   }
    \label{fig:AccelRatio}
  \end{figure}

We also ran +XY \& $-$XY simulations for planets at the middle and outer edges of the HZ. We excluded from further analysis the single situation whose mirror did not survive even without RP.  In the remaining +XY \& $-$XY simulations with RP, 
mirror survival times were typically longer for mirrors orbiting closer to their planets and around planets farther from their stars. In these situations, the RP acceleration on the mirror is less important relative to the planet's gravitational influence. Figure~\ref{fig:AccelRatio} shows survival times against the ratio between acceleration caused by RP and acceleration caused by the planet. Different colors are for different planet locations within the HZ, while different symbol shapes designate initial mirror orbit sizes. For the same planet location (symbol shading/color), mirrors orbiting closer to their planet (circles) tend to survive longest. For the same initial mirror orbit size (symbol shape), mirrors around planets farthest from their star (solid / purple) tend to survive longest. 

\section{Final Thoughts}

Mirror orbit survival in the presence of RP rarely happens naturally. Although there are trends in simulation outcomes, there is no simple predictor for which mirrors orbit long-term and which do not. 
Mirror orbits that survive long-term often have complex orbital trajectories.

\section{Acknowledgements}

We wish to thank the many UWL undergraduate students who have worked on this project since 2017, in particular Kaisa Ackerman who designed the initial python simulation interface.

\end{document}